\documentclass[conference, 10pt]{IEEEtran}
\ifCLASSINFOpdf
\else
\fi
\usepackage{subfig}
\usepackage{graphicx}
\usepackage{color}
\usepackage{placeins}
\usepackage{float}
\usepackage{hyperref}
\usepackage{tabularx,colortbl}

\usepackage{amsmath,amssymb,amstext} 
\usepackage{mathastext}
\usepackage{epsfig}

\newcommand{\revision}[1]{\textcolor{black}{#1}}

\begin{document}
%
\title{Low-Dispersive Permittivity Measurement Based on Transmitted Power Only}

\author{\IEEEauthorblockN{Seyed Hossein Mirjahanmardi}
\IEEEauthorblockA{Department of Electrical and Computer Engineering\\
University of Watreloo\\
Waterloo, Ontario, Canada\\
shmirjah@uwaterloo.ca\\}
\and
\IEEEauthorblockN{Omar M. Ramahi}
\IEEEauthorblockA{Department of Electrical and Computer Engineering\\
University of Watreloo\\
Waterloo, Ontario, Canada\\
oramahi@uwaterloo.ca}}


%


\maketitle

\begin{abstract}

This paper \revision{presents} a complex permittivity measurement \revision{method} for low-dispersive materials as a function of frequency. \revision{The introduced} method relies only on transmitted power \revision{signals which are} collected using a spectrum analyzer/power meter, removing the need for phase measurements and a vector network analyzer. This method provides a very good accuracy \revision{along} with easy and inexpensive permittivity measurements.

\end{abstract}




%
\IEEEpeerreviewmaketitle

\section{Introduction}

Permittivity measurements of materials as a function of frequency finds applications in numerous fields, including medical imaging, food sciences, and agriculture \cite{chen_varadan, martellosio}. Several techniques have been widely used to determine material permittivities, such as open-ended coaxial, free-space, cavity resonant, and waveguides. The open-ended coaxial \revision{method} covers a wide range of frequencies and provides easy measurements; however, suffers from low accuracy \cite{Seyed, meaney_openended}. Additionally, reports show that this technique cannot provide reliable results for heterogeneous medium \cite{meaney_openended}. The free-space method provides a wide-band permittivity measurement with high accuracy, but requires costly measurement setups and large sampling preparation. Cavity resonant techniques are among the highest accurate methods, but are very narrow-band and can be only applicable to low-loss materials. Waveguide techniques provide high accuracy and moderate bandwidth of operation but need cumbersome sample preparations \cite{chen_varadan}. These methods mainly rely on both amplitude and phase measurements which require vector network analyzers. There are only a few amplitude-only techniques reported in the literature that do not need phase information \cite{Hasar}. \revision{These methods, however, cannot provide high accuracy because their preciseness rely on a specific amount of loss that should occur in the sample while signals propagate from the source to the receiver. Additionally, some of them still require a vector network analyzer to measure signals' reflections.} 

In this paper, we apply our previously introduced single-frequency permittivity reconstruction method \cite{Seyed, Seyed_nondispersive} to measure the permittivity of low-dispersive materials as a function of frequency. No phase information is used in this method and only transmitted power collected at multiple frequencies are utilized to reconstruct the permittivity values. Therefore, the need for a vector network analyzer is removed. Measurement results \revision{acquired from our method} are compared with the open-ended coaxial method's, validating the fidelity of our method.

\section{Methodology}

\begin{figure}[t!] \centering
	\includegraphics[width=0.25\textwidth]{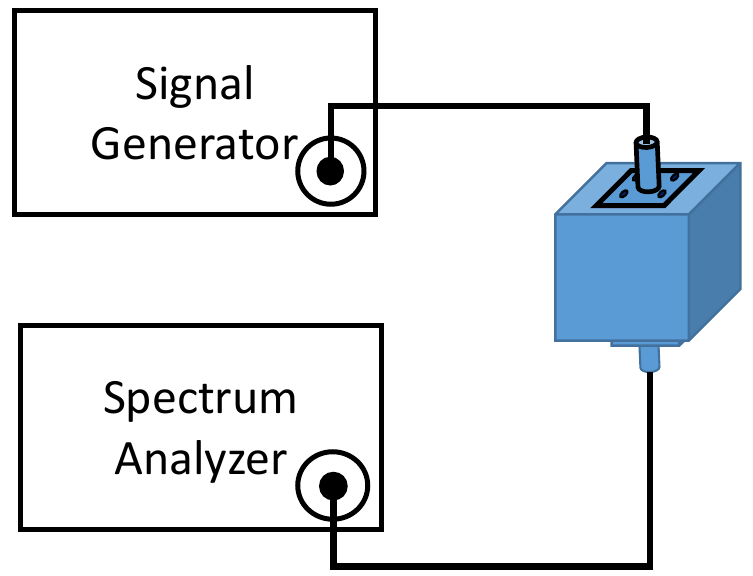}
	\caption{Coaxial line filled with \revision{a liquid} sample connected to a signal generator and a spectrum analyzer.}
	\label{fig-1}
	
\end{figure}

This section briefly discusses our power method applied to a coaxial line structure. The method can be also applied to other permittivity reconstruction setups such as the free-space one. Fig. \ref{fig-1} shows a coaxial line connected to a signal generator with an internal source resistance of $R_s$ and to a spectrum analyzer with a load resistance of $R_L$. Solving Maxwell equations for the transmitted power delivered to the spectrum analyzer analytically results in:

\begin{equation}\label{eq-power_coax}
\hat{P}_{L}(f_i, \epsilon_r', \epsilon_r'')= \frac{1}{2 R_L}{\left| \frac{V_s Z_{in} e^{- \gamma d} (\Gamma+1)}{(R_s + Z_{in})(\Gamma e^{- 2\gamma d} + 1)}  \right|^2}
\end{equation} where $V_s$, $\Gamma$, and $f_i$ are the source voltage, reflection coefficient at the load, and the frequency of excitation, respectively. $\hat{P}_{L}$ is the power calculated analytically. $Z_{in}$ is the input impedance seen from the source side \revision{toward the load}. The coaxial line has an outer radius of 4.1 mm, inner radius of 1.27 mm, and a length of 3.6 cm. The space between the inner and outer is filled with the material under test with the real permittivity of $\epsilon_r'$ and imaginary part of $\epsilon_r''$, ($\epsilon_r=\epsilon_r'-j\epsilon_r''$). The material is assumed to be non-magnetic and low-dispersive.

Equation \ref{eq-power_coax} depends non-linearly on the excitation frequency. Therefore, multiple independent equations can be generated by changing the frequency. If material under test \revision{is not highly dispersive}, its permittivity function varies smoothly versus frequency. Hence, within a finite bandwidth the permittivity can be approximately considered constant. The following equation is then defined as an objective function \revision{which compares} the transmitted power \revision{obtained analytically} with the \revision{power values obtained from measurements} at a specific frequency:

\begin{equation}\label{eq-2}
\delta (\epsilon_r', \epsilon_r'', f_i)=\left|\hat{P}_{L}(f_i)-P_L(f_i)\right|
\end{equation} where $\delta$ is the difference between measured power, $P_L$, and calculated power.

\begin{figure}[t!] \centering
	\includegraphics[width=0.25\textwidth]{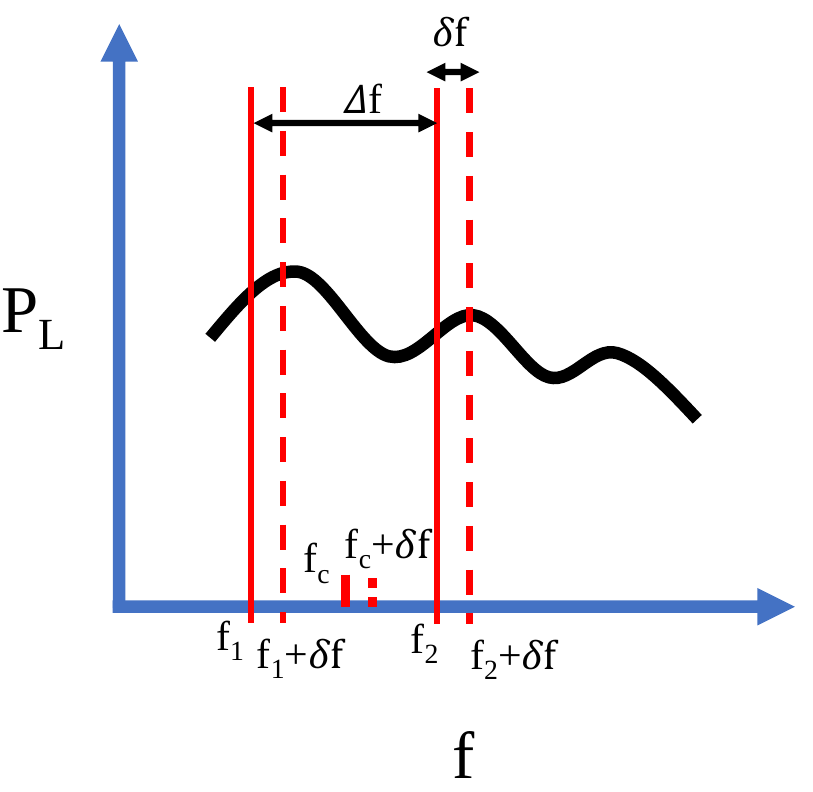}
	\caption{An example of measured power as a function of frequency. Sampling is done in a window from $f_1$ to $f_2$, shown with solid red lines, with a frequency center of $f_c$. The shifted window is shown by dashed red lines with a frequency center of $f_c+\delta f$.}
	\label{fig-2}
	
\end{figure}

Generating multiple equations based on \ref{eq-2} in a finite bandwidth $\Delta f$ creates a set of objective functions with two unknowns, $\epsilon_r'$ and $\epsilon_r''$. Multiple power samples, in this paper twenty samples, are selected linearly in this bandwidth and substituted in \ref{eq-2}. Solving these objective functions numerically to minimize $\delta$ results in determining a unique pair for $\epsilon_r'$ and $\epsilon_r''$. The obtained permittivity is assigned to the frequency located at the center of this bandwidth, $f_c$. Readers are referred to our previous works for more details on the permittivity reconstruction procedure \cite{Seyed}. Fig. \ref{fig-2} shows an example of measured power as a function of frequency. The power samples are picked up in a window from $f_1$ to $f_2$ and the permittivity is assigned to $f_c$.

This method can then be applied to determine the permittivity at other frequencies. If the bandwidth window shown in Fig. \ref{fig-2} is shifted slightly, with the amount of $\delta f$ in the frequency domain, and the permittivity reconstruction procedure is repeated, a new permittivity value is obtained and assigned to $f_c+\delta f$.

\section{Measurement Results}

\begin{figure}[t!] \centering
	\includegraphics[width=0.4\textwidth]{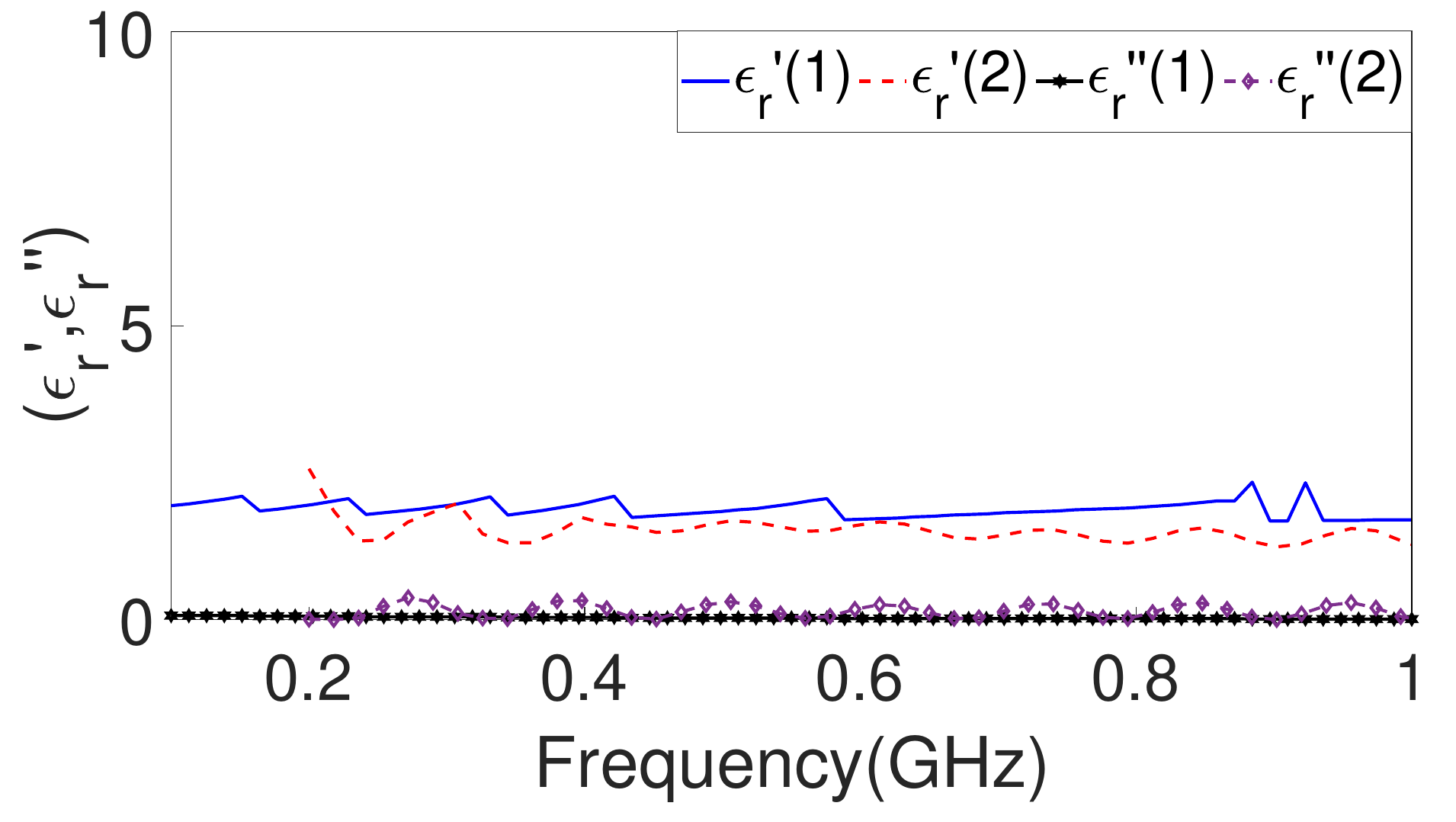}
	\caption{Real part and imaginary part of Cyclohexane liquid measured using our method, $\epsilon'(1)$, and the open-ended coaxial method, $\epsilon''(2)$.}
	\label{fig-3}
	
\end{figure}

To validate the method, a low-dispersive liquid, Cyclohexane, is tested. The coaxial line shown in Fig. \ref{fig-1} is filled with this liquid connected to a signal generator which has an output signal power level of 0 dBm. The frequency of excitation varies from 0.05 to 1.05 GHz. Twenty samples are linearly picked up with a step of 5 MHz within a bandwidth of 95 MHz and the method is applied to reconstruct the permittivity, $\epsilon_r'(1)$ and $\epsilon_r''(1)$. The window of sampling is then shifted 10 MHz in frequency and a new permittivity is measured for the new bandwidth window. Fig. \ref{fig-3} shows the result of this permittivity reconstruction for both the real and imaginary parts.

As a comparison, Cyclohexane liquid is tested with the open-ended coaxial approach. The results of this measurement are also plotted in Fig. \ref{fig-3}, $\epsilon_r'(2)$ and $\epsilon_r''(2)$. This measurement is done with a vector network analyzer. It can be seen that the results of these two methods are in close agreement. However, our previous work showed a higher accuracy of our method compared to the open-ended coaxial one \cite{Seyed}.

\section{Conclusion}

This paper introduced a permittivity measurement method for low-dispersive materials without the need for phase measurements and a vector network analyzer. Only the real part of transmitted power values collected by a spectrum analyzer are used. Comparisons with the open-ended coaxial method showed that our method provided a very good accuracy.

\bibliographystyle{IEEEtran}
\bibliography{APS_Refrences}

\end{document}